\documentclass[aps,pre,preprint,amsmath,amssymb,superscriptaddress,floatfix,showpacs]{revtex4}

\usepackage{graphicx}
\usepackage{color}

\input epsf.sty

\newlength{\TZ}
\newcommand{\BEQ}{\begin{equation}}     
\newcommand{\BEA}{\begin{eqnarray}}
\newcommand{\BD}{\begin{displaymath}}
\newcommand{\EEQ}{\end{equation}}       
\newcommand{\EEA}{\end{eqnarray}}
\newcommand{\ED}{\end{displaymath}}
\newcommand{\eps}{\varepsilon}          
\newcommand{\D}{{\rm d}}                
\newcommand{\mcEc}[1]{\multicolumn{1}{c}{#1}} 

\renewcommand{\vec}[1]{\boldsymbol{#1}} 

\begin{document}

\title{Superuniversality in phase-ordering disordered ferromagnets}

\author{Malte Henkel} 
\affiliation{Laboratoire de Physique des 
Mat\'eriaux (LPM)\footnote{Laboratoire associ\'e au CNRS UMR 7556}, 
Nancy Universit\'e, CNRS, \\ 
B.P. 239, F -- 54506 Vand{\oe}uvre l\`es Nancy Cedex, France}
\author{Michel Pleimling}
\affiliation{Department of Physics, 
Virginia Polytechnic Institute and State University,\\
Blacksburg, Virginia 24061-0435, USA}


\date{\today}

\begin{abstract}
The phase-ordering kinetics of the ferromagnetic two-dimensional Ising model 
with uniform bond disorder is investigated by intensive Monte Carlo simulations. 
Simple ageing behaviour is observed in the single-time correlator and the  
two-time responses and correlators. The dynamical exponent $z$  and the 
autocorrelation exponent $\lambda_C$ only depend on the ratio $\eps/T$, where 
$\eps$ describes the width of the distribution of the disorder, whereas a more 
complicated behaviour is found for the non-equilibrium exponent $a$ of the two-time 
response as well as for the autoresponse exponent $\lambda_R$. 
The scaling functions are observed to depend only on the 
dimensionless ratio $\eps/T$. If the length scales are measured in terms of the 
time-dependent domain size $L(t)$, the form of the scaling functions is in general 
independent of both $\eps$ and $T$. Conditions limiting the validity of this 
`superuniversality' are discussed. 
\end{abstract}
\maketitle


\section{Introduction}

Aging phenomena have become one of the paradigmatic examples which are used to
study fundamental aspects of non-equilibrium criticality, besides their practical
use in materials processing. If some physical system is brought rapidly out of 
equilibrium by a sudden change of an external control parameter 
(an often-used device is to {\em quench} the system by lowering its temperature 
rapidly from the disordered high-temperature phase to its ordered low-temperature 
phase where at least two thermodynamically stable states co-exist) 
one often finds simultaneously the
following three properties: (i) slow (i.e. non-exponential) dynamics, (ii) breaking
of time-translation invariance and (iii) dynamical scaling, which together are
said to constitute {\em ageing behaviour}. These features were first observed together 
in the mechanical properties of many polymeric materials by Struik \cite{Struik78} 
and it has since been understood that the broad characteristics 
of ageing can be found in many different types of non-equilibrium systems. 
Besides spin glasses \cite{Cugliandolo02,Vincent07}, other well-studied examples 
may be found in the phase-ordering kinetics of a ferromagnet quenched below
its critical temperature $T_c>0$ \cite{Bray94} or in granular media \cite{Dauchot07}. 

The analysis of phase-ordering as it occurs in ferromagnets quenched from an 
initially disordered state into its coexistence phase with temperature $T<T_c$ 
is particularly simple to 
formulate. The motion of the domain walls is driven by the surface 
tension between the ordered domains. The typical time-dependent length scale 
is related to the linear size of these ordered domains which grows as 
$L=L(t)\sim t^{1/z}$ where $z$ is the dynamical exponent. 
For a non-conserved order-parameter, it is well-known that $z=2$ \cite{Bray94}. Because 
of the simple algebraic scaling of the linear
domain size $L(t)$, one expects the following scaling behaviour for the
single-time correlation function
\BEQ \label{1}
C(t;\vec{r}) := \langle \phi(t,\vec{r}) \phi(t,\vec{0})\rangle = t^{-b} 
F\left( \frac{|\vec{r}|}{L(t)} \right)
\EEQ
for sufficiently large times $t\gg t_{\rm micro}$ (with a microscopic reference time
$t_{\rm micro}$ such that $L(t_{\rm micro})$ is of the order of the lattice constant). 
Similarly, for the two-time correlation and response functions (in the ageing regime, 
where the observation time $t$ and the waiting time $s$ satisfy $t,s\gg t_{\rm micro}$ 
and $t-s\gg t_{\rm micro}$):
\BEA \label{2}
C(t,s;\vec{r}) &:=& 
\langle \phi(t,\vec{r}) \phi(s,\vec{0})\rangle
=  s^{-b}
F_C\left(\frac{t}{s},\frac{|\vec{r}|}{L(s)}\right)~,~~~~\\
\label{3}
R(t,s;\vec{r})  &:=& 
\left. \frac{\delta\langle \phi(t,\vec{r})\rangle}{\delta h(s,\vec{0})}
\right|_{h=0}  = 
s^{-1-a} F_R\left(\frac{t}{s},\frac{|\vec{r}|}{L(s)}
\right)~. 
\EEA
Here $\phi(t,\vec{r})$ is the space-time-dependent order-parameter,
whereas $h(s,\vec{r})$
is the conjugate magnetic field (spatial translation-invariance of all averages 
will be assumed throughout this paper) and $a$ and $b$ are ageing exponents. The
scaling functions $F_{C,R}(y,\vec{0})\sim y^{-\lambda_{C,R}/z}$ for $y\to\infty$
which defines the autocorrelation exponent $\lambda_C$ and the autoresponse
exponent $\lambda_R$. For phase-ordering kinetics in pure systems, it is generally 
admitted that $b=0$ and simple scaling arguments show that $a=1/z$. For an initial 
high-temperature state and for pure ferromagnets, $\lambda_C=\lambda_R$ 
is independent of the known equilibrium exponents 
\cite{Bray94,Godreche02,Cugliandolo02,Chamon06,Chamon07,Mazenko06,Picone04}. 
The scenario
just described is referred to as {\em simple ageing}. The conditions for the
onset of ageing and the possible scaling forms have been carefully discussed in 
\cite{Zippold00,Andreanov06}. 

In going from phase-ordering kinetics in simple ferromagnets to glassy systems
(usually modelled by spin systems with disorder and frustration), one expects 
more complicated growth laws $L=L(t)$ which describe a cross-over from the domain
growth of essentially pure systems  (as long as $L(t)$ is small compared to 
the typical distance between disorder-created defects) to a late-time regime with a
slower growth and dominated by the defect structure. For disordered 
ferromagnets without frustration, this can be studied 
through generalisations of the Allen-Cahn equation, which attempt to describe
how the pinning of the domain walls created by the disorder should be overcome by thermal 
activation \cite{Huse85,Lai88}. In addition, it was suggested
that once the unique reference length scale is chosen to be $L(t)$, the resulting
scaling functions should become {\em superuniversal} in the sense that they should 
be {\em independent} of the disorder \cite{Fisher88}. This superuniversality has
indeed been confirmed for the single-time spin-spin correlator 
\cite{Bray91,Puri91,Hayakawa91,Iwai93,Biswal96,Aron08}. 
Furthermore, superuniversality is in qualitative
agreement with the experimental observation in several distinct polymers and metals
that the linear response to a small mechanical stress can be described in terms of an 
universal master curve which
is independent of the material studied \cite{Struik78}. 
The universal scaling functions of the {\em pure case}
can be calculated from the theory of local scale-invariance \cite{Henkel02}. Tests
in non-integrable systems include the auto- and space-time-responses in the 
two- and three-dimensional
Ising models \cite{Henkel03}, the autocorrelation function in the two-dimensional 
Ising model \cite{Henkel04}, 
the same quantities in the two-dimensional $q$-states Potts model
with $q=2,3,8$ \cite{Lorenz07} as well as the autoresponse function in several 
other cases, see \cite{Henkel07c} for a recent review. 
Therefore, with the help of superuniversality, if confirmed, the scaling functions 
describing non-equilibrium relaxation of quite complex systems
would become analytically treatable.

In this work, we shall consider a two-dimensional ferromagnetic Ising model with 
quenched bond disorder. The nearest-neighbour Hamiltonian is given by \cite{Paul04,Rieger05}
\BEQ \label{Ising_desord}
\mathcal{H} = -\sum_{(i,j)} J_{ij} \sigma_i \sigma_j, \quad
\sigma_i = \pm 1~.
\EEQ
The random variables $J_{ij}$ are uniformly distributed over
$[1-\eps/2,1+\eps/2]$ where $0 \leq \eps \leq 2 $. The model has
a second-order phase transition at a critical temperature $T_c(\eps)>0$ 
between a paramagnetic and a ferromagnetic state. It is thought that
$T_c(\eps)\approx T_c(0) \simeq 2.269\ldots$ should not depend strongly on $\eps$. 
Using heat-bath dynamics with a non-conserved order-parameter and starting
from a fully disordered initial state, phase-ordering occurs and there is
evidence which suggests that the characteristic length scale 
$L=L(t)\sim t^{1/z}$ should scale {\em algebraically} and where the
dynamical exponent $z=z(T,\eps)$ should depend continuously on the
temperature and the disorder $\eps$. Indeed, generalising the Huse-Henley heuristic
argument \cite{Huse85} by considering the case when the disorder-created energy barriers 
for the motion of the domain walls are distributed logarithmically with
respect to the domain size $L(t)$, the form \cite{Paul04,Paul05} 
\BEQ \label{z}
z =  z(T,\epsilon) \stackrel{?}{=} 2 + \epsilon/T~.
\EEQ
was proposed, where the constant $\epsilon$ parametrises the barrier height. 
Simulations of the linear domain size \cite{Paul04} seemed to confirm this, with
the empirical identification $\epsilon=\eps$ and are also consistent with
the results of field-theoretical studies in the Cardy-Ostlund 
model \cite{Schehr05}. Data from the thermoremanent magnetisation 
$M_{\rm TRM}(t,s) = h \int_0^s\!\D u\, R(t,u) = s^{-a} f_M(t/s)$ were used to estimate 
the exponent 
$a$ and, assuming $a=1/z$, also looked consistent with (\ref{z}) and $\epsilon=\eps$, 
at least for values 
for $T$ and $\eps$ for which $z$ did not become too large \cite{Henkel06a}. 
Superuniversality has 
been confirmed recently for the hull enclosed area \cite{Sicilia07}. 
On the other hand, the conclusion of a simple ageing
reached in \cite{Paul04,Paul05,Rieger05} has been questioned by more 
recent simulations for the {\it random-site} Ising model \cite{Paul07}. 
In that work, a scaling form $C(t,s) = C_{\rm st}(t-s) +C_{\rm ag}(h(t-s)/h(s))$ with
$h(u)=\exp\left(\frac{u^{1-\mu}}{1-\mu}\right)$ was considered, where $\mu$ is a fit
parameter. In the limit $\mu\to 1$, one recovers the simple ageing scenario 
described above (the stationary part $C_{\rm st}(t-s)$ merely represents an irrelevant
correction to the leading scaling behaviour). 
The case $\mu<1$ is called {\em sub-ageing}, and the case
$\mu>1$ is called {\em super-ageing}. In \cite{Paul07}, systematic deviations 
{}from the dynamical scaling of simple ageing were observed in the $2D$ 
random-site Ising model quenched to below $T_c$. A data collapse could be achieved, 
however, by allowing $\mu$ to vary and values in the range $\mu\approx 1.03 - 1.04$
were reported \cite{Paul07}. This finding was interpreted as to suggest the presence 
of a slight super-ageing effect \cite{Paul07}.

In the following, we shall present new data on the single-time and two-time 
correlations as well as on two-time response functions.
As we shall see in section~II, our data
are fully compatible with the simple ageing scenario and furthermore, looking at a
larger range of values of $z$, we find that the dynamical exponent 
$z=z(T,\eps)=z(\eps/T)$ depends on the control parameters in a more complicated way 
than suggested in eq.~(\ref{z}). These conclusions are also valid for the two-time
response function. The dependence of the various non-equilibrium exponents on both
$\eps$ and $T$ will be studied. We also show evidence that the scaling functions 
only depend on the ratio $\eps/T$ of the control parameters and finally confirm
the generic superuniversality of the scaling functions of correlation and response 
functions. However, we also find two conditions which must be satisfied for 
superuniversality to hold. 
Our conclusions are given in section~III. 

\section{Results}
The simulations are carried out as follows. 
For the integrated response
we simulated systems with ${\cal N}=300^2$ spins using the standard heat-bath algorithm. 
Prepared in an uncorrelated initial state 
corresponding to infinite temperatures,
the system is quenched to the final temperature in 
the presence of a random binary field $h_i=\pm h$
with strength $h = 0.05$, following the well-established method of Barrat \cite{Barrat98}
(using a random field avoids a bias which would drive the system rapidly out of the 
scaling regime). Turning off the random field after a waiting time $s$, the thermoremanent
magnetisation 
$M_{\rm TRM}(t,s)=\frac{1}{T{\cal N}}\sum_i \overline{\langle \sigma_i(t) h_i(s)\rangle}$ 
is measured at time $t$. 
We averaged over at least $5\cdot 10^4$ different runs with different initial 
states and different realizations of the noise. We point out that the data 
discussed in this paper are of much higher
quality than our earlier data \cite{Henkel06a} for the autoresponse.
For the autocorrelation function, we considered systems with up to
$600^2$ spins in order to avoid the appearance of finite-size 
effects for the times accessed in the simulations. The data discussed in the
following have been obtained after averaging over at least 5000 different runs
with different random numbers. Our main focus was on $\eps=0.5$, $\eps=1$, 
and $\eps=2$ where we considered for every case at 
least four different temperatures. In addition, some runs where also
done for other values of $\eps$. The total study took 
approximately $2 \cdot 10^5$ CPU hours on Virginia Tech's System X supercomputer 
composed of Dual 2.3 GHz PowerPC 970FX processors.

\begin{figure*}[t]
\begin{center}
     \includegraphics[width=6.25in,angle=0,clip=]{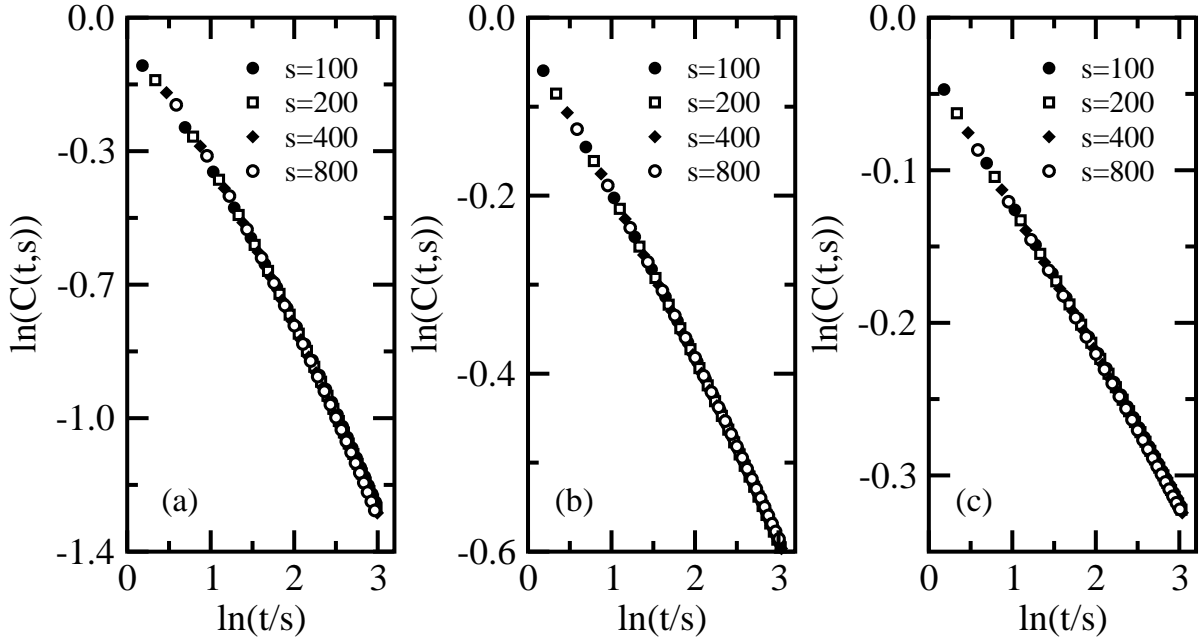}
\end{center}
\caption[Fig 1]{Scaling of the autocorrelation $C(t,s)$ for (a) $\eps=0.5$, $T=1$,
(b) $\eps=1$, $T=0.4$ and (c) $\eps=2$, $T=0.4$, for several values of $s$. 
Here and in the following error bars are smaller than the sizes of the symbols, 
unless explicitly stated otherwise.
\label{Fig1}
}
\end{figure*}

Our first question is about the scaling form to be used. 
In figure~\ref{Fig1}, we show data for the autocorrelation 
$C(t,s) = {\cal N}^{-1} \sum_i \langle \sigma_i(t)\sigma_i(s)\rangle$, plotted
over against $t/s$, for several typical values of $\eps$ and $T$. A nice data
collapse is seen, which is fully consistent with simple ageing. Our scaling plots also imply 
that the exponent $b=0$, analogously to what is found in the
phase-ordering of pure systems. In \cite{Paul07}, a `super-ageing' scaling form
$C(t,s) = {\cal C}\bigl( \exp\bigl[ \frac{(t-s)^{1-\mu}-s^{1-\mu}}{1-\mu}\bigr]\bigr)$ 
was considered for the random-site Ising model where the exponent $\mu>1$ is fitted 
to the data. Simple ageing is recovered in the 
$\mu\to 1$ limit. However, the values of $\mu\approx 1.03$ reported in \cite{Paul07} 
are so close to 
unity that a careful study on possible finite-time corrections to scaling 
appears to be required before such a conclusion could be 
accepted \cite{footnote1}.
For all $\eps<2$, our data for the random-bond model show no hint for a `super-ageing' 
behaviour \cite{footnote2}, in contrast to the findings in \cite{Paul07}. 
Note that the observed scaling form of simple ageing would be incompatible with 
a non-power-law form of $L(t)$ in the range of times considered. 
We also remark that for $\eps=2$ and larger temperatures finite-time corrections
to simple ageing are observed, see \cite{Henkel06a,Henkel07a}.

\begin{figure*}[t]
\begin{center}
     \includegraphics[width=4.75in,angle=0,clip=]{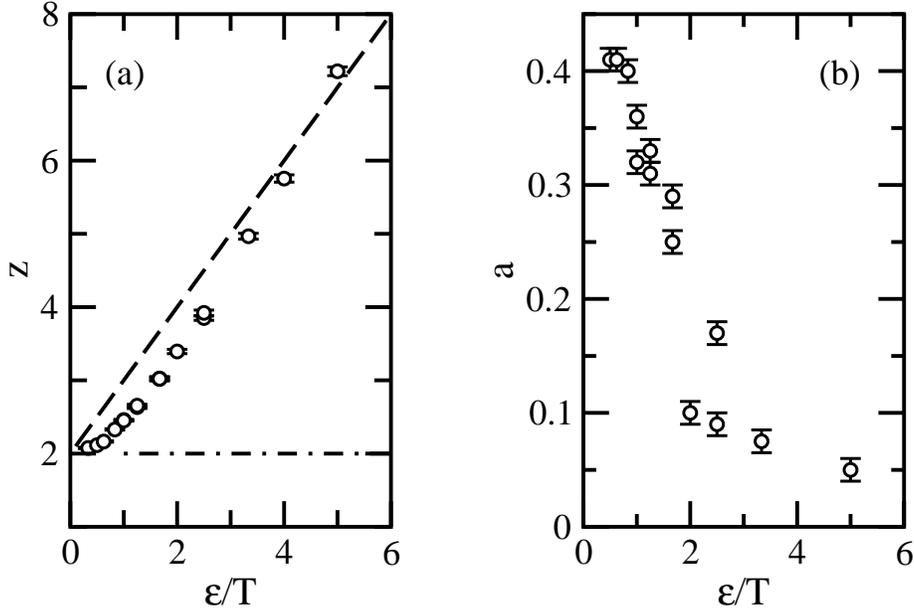}
   \end{center}
\caption[Fig 2]{Left panel: Dynamical exponent $z$, 
determined from the condition $C(t,L(t))\stackrel{!}{=}\frac{1}{2}$, as a function 
of $\eps/T$. The dashed line corresponds to the equation (\ref{z}). 
Right panel: ageing exponent $a$, as determined from the scaling of $M_{\rm TRM}(t,s)$.
In some cases we have more than one data point for a given value of $\eps/T$. These data points
correspond to different values of $\eps$ and $T$ with $\eps/T$ constant, see 
table~\ref{Tabelle1}, showing that $a$ is not simply a function of the ratio $\eps/T$
but that it depends in a more complicated way on both $\eps$ and $T$.
\label{Fig2}
}
\end{figure*}

\begin{table}
\begin{center}
\begin{tabular}{|cc|cllc|} \hline \hline
$\eps$ & $T$ & $z$ & \multicolumn{1}{c}{$a~$} & \multicolumn{1}{c}{$\lambda_C/z~$} & $\lambda_R/z$ \\ \hline
~$0.5$~& ~$1.5$~ & ~$2.08(1)$~ & \mcEc{$-~~$}    & ~$0.59(1)$  &  $-$        \\ 
~$0.5$~& ~$1.0$~ & ~$2.11(1)$~ & ~$0.41(1)~ $  & ~$0.565(10)$~ & ~$0.61(2)$~ \\ 
~$0.5$~& ~$0.8$~ & ~$2.16(1)$~ & ~$0.41(1)~ $  & ~$0.56(1)$~   & ~$0.61(3)$~ \\ 
~$0.5$~& ~$0.6$~ & ~$2.33(1)$~ & ~$0.40(1)~ $  & ~$0.54(1)$~   & ~$0.60(2)$~ \\ 
~$0.5$~& ~$0.5$~ & ~$2.46(1)$~ & ~$0.36(1)~ $  & ~$0.48(1)$~   & ~$0.55(2)$~ \\ 
~$0.5$~& ~$0.4$~ & ~$2.64(2)$~ & ~$0.33(1)~ $  & ~$0.46(1)$~   & ~$0.50(2)$~ \\ 
~$0.5$~& ~$0.3$~ & ~$3.02(2)$~ & ~$0.29(1)~ $  & ~$0.385(10)$~ & ~$0.46(2)$~ \\ 
~$1.0$~& ~$1.0$~ & ~$2.45(1)$~ & ~$0.32(1)~ $  & ~$0.49(1)$~   & ~$0.51(2)$~ \\ 
~$1.0$~& ~$0.8$~ & ~$2.65(2)$~ & ~$0.31(1)~ $  & ~$0.445(10)$~ & ~$0.50(2)$~ \\ 
~$1.0$~& ~$0.6$~ & ~$3.02(2)$~ & ~$0.25(1)~ $  & ~$0.38(1)$~   & ~$0.43(2)$~ \\ 
~$1.0$~& ~$0.4$~ & ~$3.85(3)$~ & ~$0.17(1)~ $  & ~$0.29(1)$~   & ~$0.34(1)$~ \\ 
~$1.5$~& ~$0.9$~ & ~$3.02(3)$~ & \mcEc{$-~~$}  & ~$0.375(10)$~ &  $-$        \\ 
~$2.0$~& ~$1.0$~ & ~$3.39(3)$~ & ~$0.10(1)~ $  & ~$0.315(10)$~ & ~$0.35(2)$~ \\ 
~$2.0$~& ~$0.8$~ & ~$3.92(4)$~ & ~$0.09(1)~ $  & ~$0.270(5)$~  & ~$0.32(2)$~ \\ 
~$2.0$~& ~$0.6$~ & ~$4.97(4)$~ & ~$0.075(10)~ $ & ~$0.217(3)$~ & ~$0.28(1)$~ \\ 
~$2.0$~& ~$0.5$~ & ~$5.76(5)$~ & \mcEc{$-~~$}  & ~$0.189(3)$~  &  $-$        \\ 
~$2.0$~& ~$0.4$~ & ~$7.22(6)$~ & ~$0.05(1)~ $  & ~$0.155(3)$~  & $0.21(1)$~  \\ 
\hline \hline
\end{tabular}
\end{center}
\caption{Dynamical exponent $z$, non-equilibrium response exponent $a$, the
autocorrelation exponent $\lambda_C/z$ and the autoresponse exponent $\lambda_R/z$ 
for different values of $\eps$ and $T$. 
\label{Tabelle1}
}
\end{table}

Having in this way checked that the relevant length scale $L=L(t)$ should 
indeed scale algebraically 
with time, we next determined the dynamical exponent
$z=z(T,\eps)$ from the criterion \cite{Sicilia07} involving the single-time correlator
\BEQ \label{gl6}
C(t,L(t)) \stackrel{!}{=} \frac{1}{2} \;\; , \;\; L(t) \sim t^{1/z(T,\eps)}.
\EEQ
The results for $z$ are shown in figure~\ref{Fig2} and listed in 
table~\ref{Tabelle1}. First, we observe that
the values of $z$ obtained in fact only depend on the dimensionless ratio $\eps/T$,
to within our numerical accuracy.
Second, we see that the function $z=z(\eps/T)$ is {\em non}-linear and only becomes an
approximately linear function in a relatively small region of values of $z$. 
We are confident that our results are more reliable than earlier ones since
they do not just describe the scaling of a single quantity, but rather will be
needed for the correct scaling description of several other observables, 
as we shall show below. We stress
that only the values of $z$ as given in table~\ref{Tabelle1} will lead to a good
scaling according to simple ageing without having to consider possible corrections to
scaling.

\begin{figure*}[t]
\begin{center}
     \includegraphics[width=6.25in,angle=0,clip=]{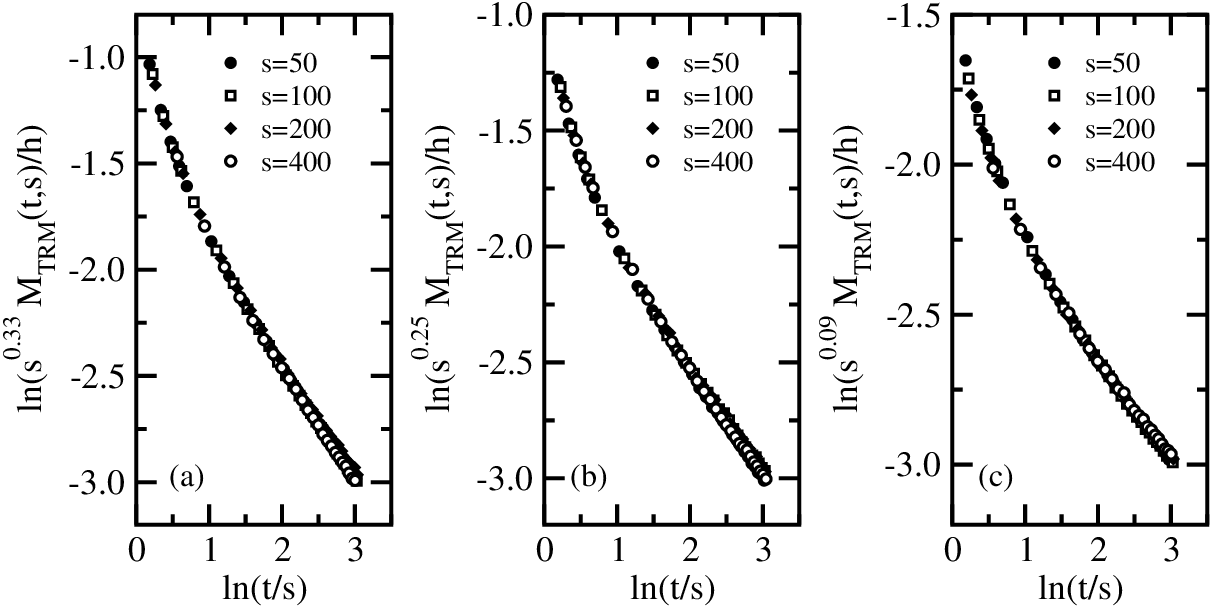}
   \end{center}
\caption[Fig 3]{Scaling of the thermoremanent magnetisation $M_{\rm TRM}(t,s)$ for
several values of $s$, with (a) $\eps=0.5$, $T=0.4$, (b) $\eps=1$, $T=0.6$ and (c)
$\eps=2$, $T=0.8$. 
\label{Fig3}
}
\end{figure*}

In the same way, in figure~\ref{Fig3} we show data for the scaling of the
thermoremanent magnetisation, expected to be of the form 
\begin{equation} \label{scal_M}
M_{\rm TRM}(t,s) = s^{-a} f_M(t/s)~. 
\end{equation}
The exponent $a$ is obtained in the usual way \cite{Henkel05} by plotting
the thermoremanent magnetisation as a function of the waiting time $s$ for
fixed values of the ratio $t/s$. The resulting power-law decay, see eq.~(\ref{scal_M}),
then yields the value of the exponent $a$. 
The numerical values and error bars given in table~\ref{Tabelle1}
and shown in figure~\ref{Fig2} 
are obtained after averaging over the values obtained for
five different values of $t/s$, namely 3, 5, 7, 10, and 15.
Looking at these values (see figure~\ref{Fig2}), we observe that the estimates 
$a=a(T,\eps)$ scatter considerably more than those for $z$. We consider 
this scatter to be large enough to conclude that $a$ cannot be reduced to a 
function of the single variable $\eps/T$. 
Furthermore, considering in detail the numerical values from table~\ref{Tabelle1},
we see that the relation $a=1/z$, known from the
phase-ordering of pure ferromagnets \cite{Cugliandolo02,Bouchaud00,Henkel03a}, 
is no longer valid. 

In order to understand this finding, let us briefly reconsider how the relation
$a=1/z$ may be derived for pure ferromagnets. Consider a pure ferromagnet in an external 
oscillating magnetic field of angular frequency $\omega$. The dissipative
part of the linear response is given as the imaginary part of the dynamic
susceptibility and reads (see e.g. \cite{Cugliandolo02})
\BEQ \label{glchi1}
\chi''(\omega,t) = \int_0^{t} \!\D u \, R(t,u) \sin\bigl(\omega(t-u)\bigr) 
= \chi_1''(\omega) + t^{-a} \chi_2''(\omega t) + \ldots
\EEQ
where the last relation follows from the usually assumed scaling (\ref{3}) of the 
autoresponse function $R(t,s)$. 
On the other hand, motivated from the physical picture that
the dynamics in phase-ordering should only come from the motion of the domain walls
between the ordered domains, one would expect to find \cite{Bouchaud00} 
\BEQ \label{glchi2}
\chi''(\omega,t) = \chi_{\rm st}''(\omega) + L(t)^{-1} \chi_{\rm age}''(\omega t) +\ldots
\EEQ
{}from which one may identify the stationary and the ageing part with the terms 
in eq.~(\ref{glchi1}) coming 
{}from the scaling analysis. Since only the domain boundaries contribute to the
dynamics, the leading time-dependent part should be proportional to the 
surface area of the domain 
divided by the total volume, hence to $L(t)^{d-1}/L(t)^d = 1/L(t)$
which accounts for the factor $1/L(t)$ in (\ref{glchi2}). Comparison of 
eqs.~(\ref{glchi1},\ref{glchi2}), together with $L(t)\sim t^{1/z}$, then gives $a=1/z$. 

Our empirical observation that $az < 1$ suggests that the above argument should 
no longer apply to {\it random} ferromagnets. Since (\ref{glchi1}) only depends
on the dynamical scaling assumption (\ref{3}), and given that our numerical results 
appear to be compatible with it, we expect that (\ref{glchi1}) should remain 
valid for disordered ferromagnets. Since also for 
disordered ferromagnets, the contribution to the ageing behaviour should come 
from the boundary region between ordered domains (this is also suggested by 
looking at the microscopic spin configurations, 
see e.g. \cite{Henkel07a}), we expect it to be proportional to $N_d(L)/N_b(L)$ 
where $N_{b,d}(L)$ denote the number of mobile spins in the bulk and at the 
domain boundaries, respectively. 
While one should still have $N_b(L)\sim L^d$, disorder may cause the domain 
boundary to become fractal and, hence, $N_d(L)\sim L^{d_{\rm f}}$ with
$d_{\rm f}$ the fractal dimension (and $d_{\rm f}=d-1$ for the pure case). 
Then eq.~(\ref{glchi2}) would be replaced by
\BEQ \label{glchi3}
\chi''(\omega,t) = \chi_{\rm st}''(\omega) + 
L(t)^{-(d-d_{\rm f})} \chi_{\rm age}''(\omega t) +\ldots
\EEQ
and comparison with (\ref{glchi1}) would now imply
\BEQ \label{az}
a = \frac{d - d_{\rm f}}{z}~.
\EEQ
Our empirical results (table~\ref{Tabelle1}) imply that $d_{\rm f}> d-1$, that is,
the disorder should modify the domain boundaries into fractal curves. From eq.~(\ref{az}), 
since $a$ depends on the dynamical exponent $z=z(\eps/T)$ 
as well as the fractal dimension $d_{\rm f}$, it may appear more natural that 
$a$ {\em cannot} be written as a function of the single variable $\eps/T$. 

\begin{figure*}[t]
\begin{center}
     \includegraphics[width=6.25in,angle=0,clip=]{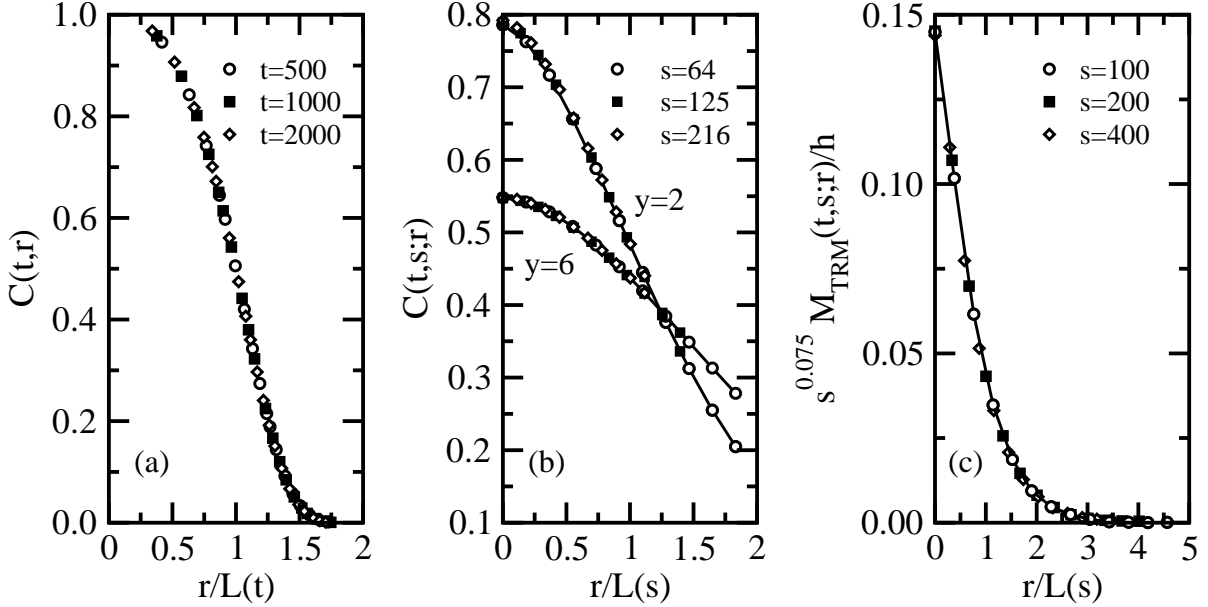}
   \end{center}
\caption[Fig 4]{(a) Scaling behaviour of the single-time correlator $C(t,\vec{r})$ 
for $\eps=1$ and $T=0.8$ and several 
values of $t$, as a function of the scaling variable $r/L(t)$. (b) Scaling of
the space-time-dependent correlator $C(t,s;\vec{r})$ for $\eps=0.5$ and $T=0.5$ for 
several values of $s$ and two fixed values of $y=t/s$, as a function of the scaling
variable $r/L(s)$. (c) Scaling of
the space-time-dependent thermoremanent magnetisation $M_{TRM}(t,s;\vec{r})$ for 
$\eps=2$ and $T=0.6$ for several values of $s$,
with $y=t/s=5$. The lines are guides to the eye. 
\label{Fig4}
}
\end{figure*}

In figure~\ref{Fig4} we show the scaling behaviour of the space- and time-dependent
correlation and response functions for various values of the waiting time $s$. 
For the selected typical values of $\eps$ and $T$, a simple ageing
behaviour is observed, in agreement with the observed scaling behaviour of the 
autocorrelation and 
of the autoresponse. We have found completely analogous
results for all other values of $\eps$ and of $T$ which we considered. 
In fact, the scaling is much cleaner than for the $\vec{r}=0$
quantities and for none of the studied cases a sizeable correction to scaling 
could be
identified. Obviously, space- and time-dependent quantities are much better suited 
for an investigation of the scaling forms than quantities that only depend on time.
Similar conclusions have recently been drawn from a study of nonequilibrium growth 
models \cite{Rot06}.

Using the scaling forms (\ref{2},\ref{3}) in the limit of large $y=t/s$ for the
autocorrelation and autoresponse functions (where $\vec{r}=0$), we have also 
extracted the exponents $\lambda_C/z$ and $\lambda_R/z$ and list our results in
table~\ref{Tabelle1}. In contrast to the pure case, where for fully disordered
initial conditions one may show that $\lambda_C=\lambda_R$ \cite{Bray94,Picone04},
the values of the autocorrelation exponent $\lambda_C$ are different from those of
the autoresponse exponent $\lambda_R$. In particular, we find that within our
numerical accuracy, $\lambda_C/z$ is a function of the single variable $\eps/T$,
at least for $\eps<2$, while $\lambda_R/z$ cannot be expressed in this way. Our
data suggest that $\lambda_R\geq \lambda_C$ and they are consistent
with the rigorous Yeung-Rao-Desai inequality $\lambda_C\geq d/2$ \cite{Yeung96}. 
Furthermore, we observe that $\lambda_R/z-a$ should be practically
constant (again for $\eps<2$ and with a value in the range $\approx 0.17-0.20$). 

In conclusion, our data are clearly consistent with 
simple ageing of the single-time and
two-time correlation functions, as well as for the thermoremanent magnetisation, 
and fully confirm the anticipated scaling forms 
(\ref{1},\ref{2},\ref{3}) with $L(t)\sim t^{1/z}$.

\begin{figure}[t]
\begin{center}
     \includegraphics[width=6.25in,angle=0,clip=]{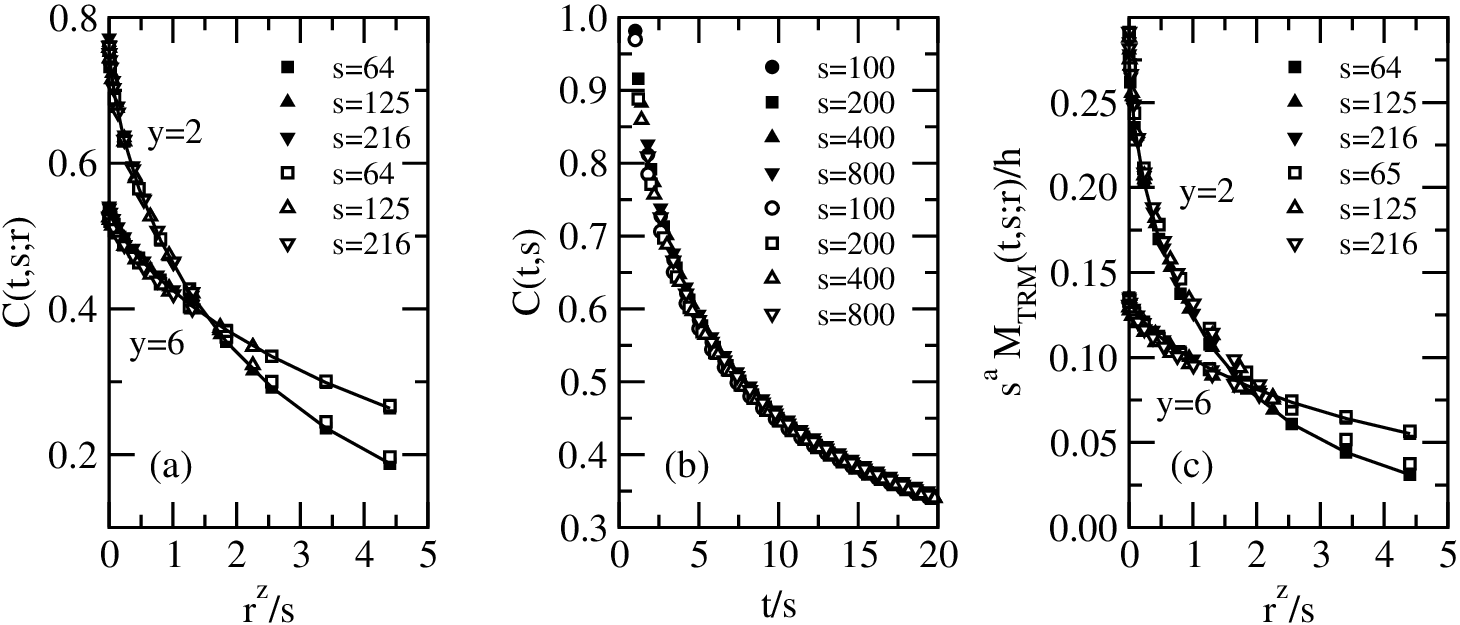}
   \end{center}
\caption[Fig 5]{(a) Space-time correlator $C(t,s;\vec{r})$ for $\eps=T=1$ (filled symbols) 
and $\eps=T=0.5$ (open symbols) for several waiting times $s$ and the 
values $y=t/s=2$ and $y=t/s=6$. The full curves are guides to the eye. 
(b) Autocorrelator $C(t,s)$ for $\eps=T=1$ (filled symbols) and $\eps=T=0.5$ 
(open symbols) for several waiting times $s$. (c) Scaled space-time 
thermoremanent magnetisation for $\eps=T=1$ (filled symbols) (with $a=0.32$) 
and $\eps=T=0.5$ (open symbols) (with $a=0.36$) for several waiting times $s$ and the 
values $y=t/s=2$ and $y=t/s=6$. The full curves are guides to the eye. 
\label{Fig5}
}
\end{figure}

Next, we shall compare the form of the scaling functions, for several values of the
control parameters $\eps$ and $T$. Since the dynamical exponent $z=z(\eps/T)$, see
figure~\ref{Fig2}, one might expect that the scaling functions themselves should
only depend on the ratio $\eps/T$, rather than on $\eps$ and $T$ separately. 
In figure~\ref{Fig5}, we test this idea by comparing data for $\eps=T=1$ with
those for $\eps=T=0.5$, for the three cases of 
(a) the space-time-dependent correlation 
$C(t,s;\vec{r})$, (b) the autocorrelation $C(t,s)$ and (c) the
space-time-dependent thermoremanent magnetisation $M_{\rm TRM}(t,s;\vec{r}) = h 
\int_0^s\!\D u\, R(t,u;\vec{r})$. In all cases, there is a clear scaling behaviour
consistent with simple ageing \cite{footnote3}
and the scaling functions nicely superpose (for the integrated response, the
data for $\eps=T=0.5$ were multiplied by 1.17 in order to take into account
the well-known presence of the non-universal numerical prefactor). 
This result, namely that the form
of the scaling function only depends on the ratio $\eps/T$, goes beyond the
standard scaling form (\ref{1},\ref{2},\ref{3}), yet it does not require to re-scale
the length by the typical domain size $L(t)$, as it would be required for a 
test of superuniversality. More systematic tests of this result would be welcome. 
We point out that the findings of figure~\ref{Fig5} are consistent with our
earlier observation that $\lambda_C=\lambda_C(\eps/T)$. The more complicated
dependence of $\lambda_R$ on both $\eps$ and $T$ would only appear if in plots
such as figure~\ref{Fig5}c one would concentrate on the region $|\vec{r}|\approx 0$.

\begin{figure}[t]
\begin{center}
     \includegraphics[width=6.5in,angle=0,clip=]{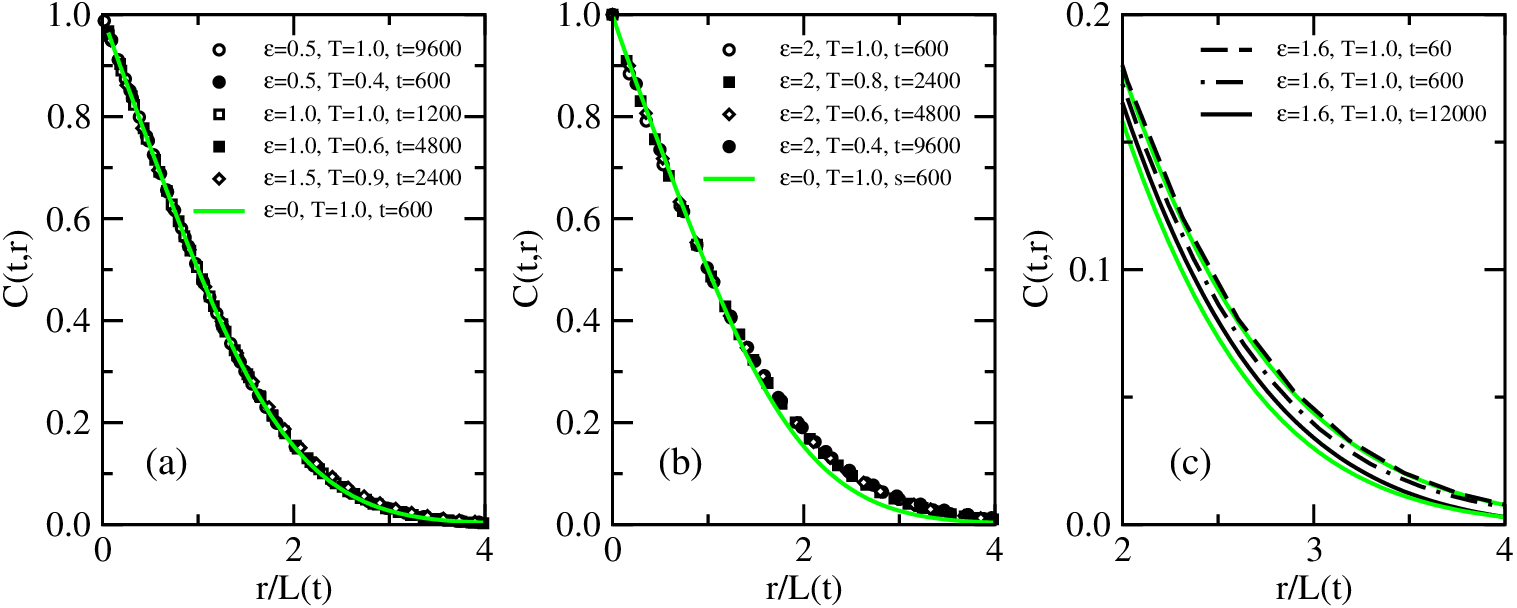}
\end{center}
\caption[Fig 6]{(Colour online) Test of the superuniversality for the single-time 
correlation function. In panel (a), data for $\eps=0, 0.5, 1, 1.5$ are shown 
to fall onto the same rescaled curve. In panel (b), we show that the data 
for $\eps=2$ describe a
curve different from the one of panel (a). The green curve is
the scaling function of the pure system.
In panel (c), data for $\eps=1.6$ and different times are compared with the
scaling curves as measured for $\eps=0$ (lower green line) and $\eps=2$
(upper green line). 
\label{Fig6}
}
\end{figure}

After these preparations, we are ready for a test of superuniversality. 
Superuniversality of the single-time correlator $C(t,\vec{r})$ is tested in figure~\ref{Fig6}. 
First, we show in figure~\ref{Fig6}a data for several
values of $0\leq \eps<2$ and $T$, where the values of $z$ may be read off from 
figure~\ref{Fig2} and table \ref{Tabelle1}. 
The times were chosen sufficiently large such that a clear scaling behaviour 
has set in. Using the typical length scale $L(t)$ as determined
earlier from eq.~(\ref{gl6}), we see that indeed all data, including the one for the case 
without disorder, collapse nicely onto a single curve, within the
numerical accuracy. This is a clear confirmation of superuniversality, very much
in agreement with earlier studies 
\cite{Bray91,Puri91,Hayakawa91,Iwai93,Biswal96,Sicilia07,Aron08}. However, 
when considering the case $\eps=2$, which is shown in figure~\ref{Fig6}b, a 
different picture emerges. Clearly, the scaling curves for $C(t,\vec{r})$ as obtained for
several values of $T$ again collapse onto each other, but, as the comparison with
the scaling function of the pure case shows, the scaling function is no longer
the same as the one of the pure case. 
Therefore, if taken at face value, the case $\eps=2$ might represent
a distinct superuniversality class. In order to get a better understanding on this point, we
show in figure~\ref{Fig6}c data for a relatively large value of $\eps$. One sees that
for moderately large times, the data are very close to the curve found for $\eps=2$, but when
the time $t$ is made very large, a cross-over to the curve of the pure case $\eps=0$ 
is observed. Qualitatively, the cross-over time $t_{\times}$ increases when $\eps\to 2$ and
becomes so large that a cross-over is no longer detectable for the times (and the lattice sizes,
which must be increased to large times in order to avoid finite-size effects) reachable with
our numerical methods. We did not see any sign for a cross-over in our data with $\eps=2$, 
but purely numerical techniques cannot distinguish between a very large and an infinite
cross-over time $t_{\times}$. 

\begin{figure}[t]
\begin{center}
     \includegraphics[width=5.00in,angle=0,clip=]{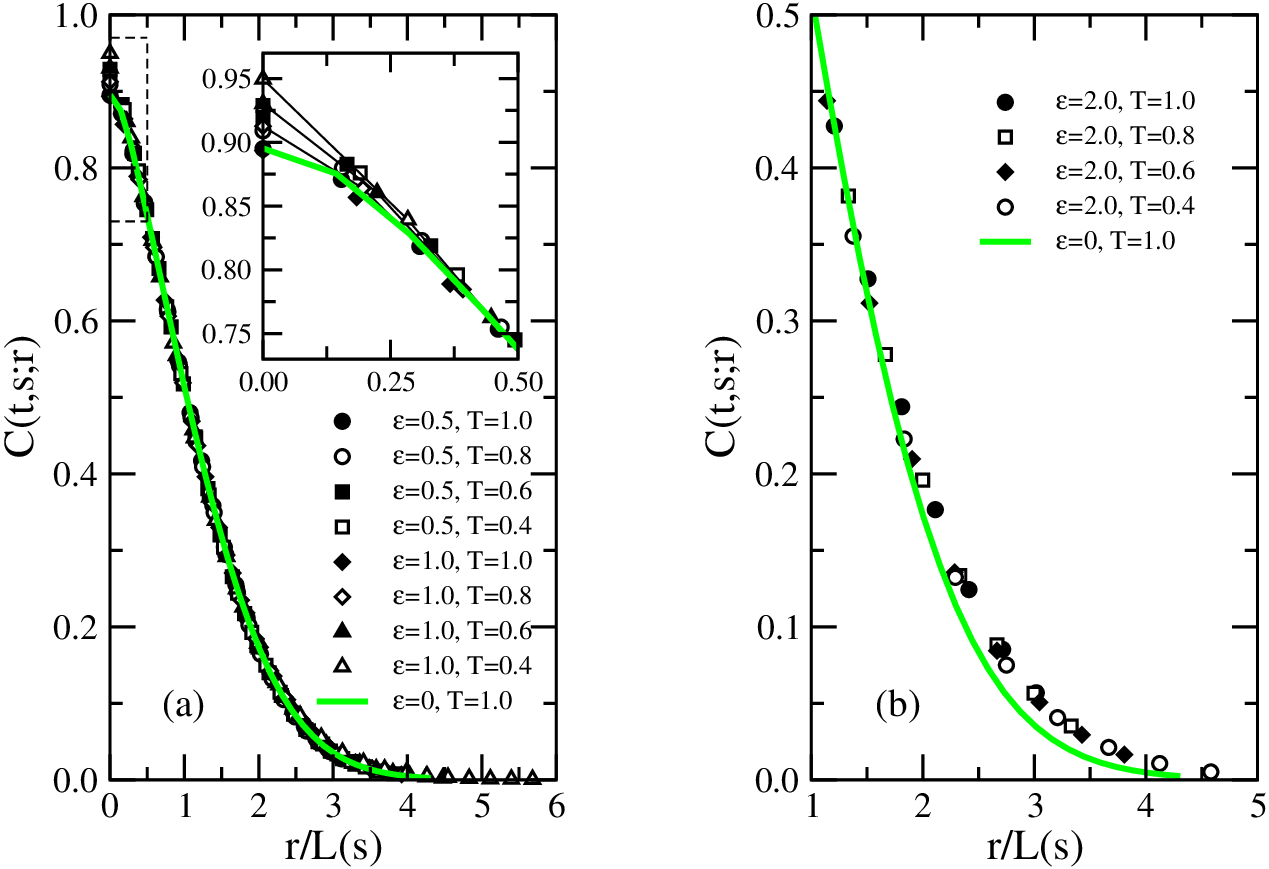}
   \end{center}
\caption[Fig 7]{(Colour online) Test of the superuniversality of the two-time 
correlation function, 
with $y=t/s=4$ and analogous to Figure {\protect \ref{Fig6}}. 
The data shown are for $s=100$. 
The green curves show the data of the pure system.
The inset in panel (a) is a magnification of the area within the dashed box. 
The black lines are guides to the eye.
\label{Fig7}
}
\end{figure}

\begin{figure}[bt]
\begin{center}
     \includegraphics[width=5.00in,angle=0,clip=]{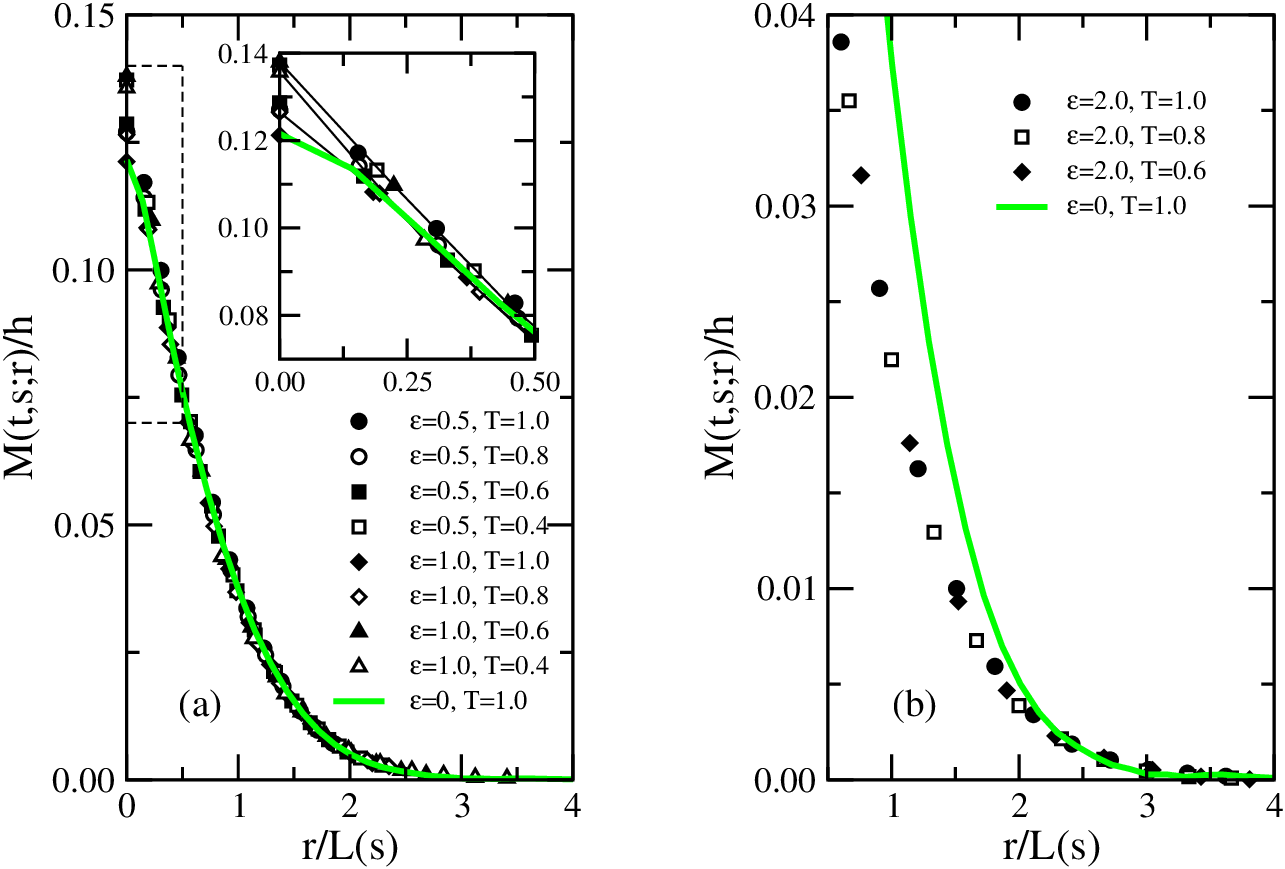}
   \end{center}
\caption[Fig 8]{(Colour online) Test of the superuniversality of the two-time 
thermoremanent magnetisation, with $y=t/s=4$ and analogous to Figure {\protect \ref{Fig6}}. 
The data shown are for $s=100$. 
In order to compare the data for different $\eps$ and $T$,
we have to take care of the fact that the response contains a non-universal multiplicative
factor. The data have been multiplied by a constant in such a way that the different
curves coincide with the curve for $\eps = 0$ and $T=1$ for large values of $\vec{r}/L(s)$.
The green curves show the data of the pure system.
The inset in panel (a) is a magnification of the area within the dashed box.
The black lines are guides to the eye.
\label{Fig8}
}
\end{figure}

We extend the test of superuniversality to the case of space-time-dependent
two-time correlators in figure~\ref{Fig7} and similarly for the space-time-dependent 
two-time response in figure~\ref{Fig8}. Qualitatively, we arrive at essentially the
same conclusion as for the single-time correlator. 
In both cases, panel (a) demonstrates a superuniversal behaviour for values 
of $0\leq \eps<2$. We show here our data for $y=t/s=4$, but
the same behaviour is observed for other values of $t/s$ accessed in this study, namely
$2 \leq t/s \leq 10$. However, closer inspection also shows that superuniversality is
no longer true for relatively small spatial distances $|\vec{r}|/L(s) \lesssim 0.5$.
This is shown in the insets of the panels (a) in figures \ref{Fig7} and \ref{Fig8}. 
Indeed, systematic deviations are observed for small spatial distances, 
the largest deviations
being observed for the autocorrelation and autoresponse functions 
with $|\vec{r}| =0$. The value of $|\vec{r}|/L(s)$ where the deviations set in 
seems to depend slightly on the 
value of $s$, but a larger range of $s$ values than accessed in the present 
study is needed for a more quantitative discussion of this point.
That means that although dynamical scaling
does hold true even down to the autocorrelators and autoresponses, correlators and
responses taken over a spatial distance of at least a typical cluster size $L(t)$
show yet a larger degree of universality. 

This requirement appears to be consistent with the known numerical values of
the exponents $\lambda_C/z$ and $\lambda_R/z$ as listed in table~\ref{Tabelle1}. 
Superuniversality at $\vec{r}=\vec{0}$ 
would have required that their values should have been equal to
those of the pure case $\eps=0$, but we have rather seen that they depend on $\eps$ and $T$.  

On the other hand, the case $\eps=2$ again stands apart, as we illustrate in the
panels (b) in both figure~\ref{Fig7} and~\ref{Fig8}. Comparing the data for $\eps=2$
with the scaling functions found for $0\leq \eps<2$ (see panels (a)), we find
small but systematic deviations. We stress that although these deviations are not very large, 
they are well outside the error bars of our high-quality data. 
As for the single-time 
correlator, it remains a possibility that for enormous times there might occur a cross-over
to the scaling function of the pure case, but the relevant cross-over time $t_{\times}$ 
is far larger than the time scales reached by our simulation.

\section{Discussion}

In this work, we have studied the non-equilibrium scaling behaviour of a disordered
Ising model without frustration, in an attempt to appreciate better the role
of disorder by considering its effects in a system which is no longer identical
to a pure system but which yet does not show the full complexity of a spin glass.
Our conclusions are as follows:
\begin{enumerate}
\item When quenched to a temperature $T<T_c$ from a totally disordered state, the
two-dimensional bond-disordered Ising model undergoes phase-ordering kinetics. 
The typical length scale $L(t)\sim t^{1/z}$ of the ordered domains 
scales algebraically with time, where the dynamical exponent
$z=z(\eps/T)$ depends continuously on the dimensionless ratio $\eps/T$ of the 
control parameters. 

Quantitatively, this dependence can be read off from figure~\ref{Fig2}, and does not
agree with earlier proposals of a linear relation such as in eq.~(\ref{z}). 
We have seen that this value of $z$ correctly describes the dynamical scaling of not
only single-time correlators, but of the two-time correlators and responses as well.

 
\item Our data are completely compatible with simple ageing. 
\item While the non-equilibrium exponent $b=0$ of the correlation function is 
un-modified with respect to pure phase-ordering systems, the non-equilibrium exponent 
$a$ which describes the scaling of the response function is no longer
simply related to the dynamical exponent $z$, see table~\ref{Tabelle1} and figure~\ref{Fig2}. 

We propose to account for this finding in terms of a postulated fractal
structure of the domain walls, which has led us to eq.~(\ref{az}). Further tests
of this idea would be welcome. 

The autocorrelation and autoresponse exponents $\lambda_C=\lambda_C(\eps/T)$ and 
$\lambda_R=\lambda_R(T,\eps)$ are distinct from each other, in contrast to the pure case. 
\item Our data suggest that the form of the scaling functions only depends on the
dimensionless ratio $\eps/T$. It remains to be seen to what extent this observation 
can be extended to different systems. 
\item In general, our data appear to confirm the superuniversality hypothesis, 
that is when all length scales are expressed in terms of $L(t)$, the
form of the scaling function is independent of both the disorder $\eps$ and the
temperature $T$.

However, we have also found two important qualifications: 
\begin{enumerate} 
\item Superuniversality does not hold for sufficiently small spatial distances 
$|\vec{r}|/L(t)\lesssim 0.5$. 
\item For $\eps=2$, although we 
find throughout a similar data collapse, the
form of the scaling functions no longer co\"{\i}ncide with the ones of the pure case.
It is not understood whether the data presented here 
should be viewed as giving evidence for
a distinct superuniversality class or else if there is a cross-over 
to the scaling functions
of the pure scale at time scales much larger than the ones reached in our study. 
\end{enumerate} 
A better understanding of superuniversality will require an explanation of these
conditions. 
\end{enumerate}

What can these findings tell us on the behaviour of real materials~? Indeed, it has 
been shown recently, in a comparative study of the three-dimensional 
random-{\em field} Ising 
model and the three-dimensional Edwards-Anderson spin glass \cite{Aron08}, 
that superuniversality
is apparently satisfied in the former case (in which the disorder is `weak' such that
the ground state is still ferromagnetically ordered) while in the latter 
it is not \cite{frust}. 
Our own result is in qualitative agreement with this, but it raises the question
how to explain the celebrated universality of the scaling functions for the linear response 
found in largely different materials \cite{Struik78}.

\newpage 
\noindent {\bf Acknowledgements} \\

We thank Leticia Cugliandolo for useful discussions. 
The simulations have been done on Virginia Tech's System X. 
MH thanks H. Park for kind hospitality at the KIAS Seoul, where the writing-up 
of this work was finished.



\begin{thebibliography}{99}
\bibitem{footnote1}
Indeed, if we take $\eps=2$, then for $T\gtrsim 0.6$ notable finite-time 
corrections to simple ageing are seen, which may be accounted for by 
including a finite-time correction term \cite{Henkel06a,Henkel07a}
\begin{displaymath}
C(t,s) = F_C(t/s) - s^{-b'} G_C(t/s)
\end{displaymath}
and a new correction exponent $b'>0$.
\bibitem{footnote2}
If one uses a scaling form $C(t,s)={\cal C}\bigl((t-s) s^{-\mu}\bigr)$, 
it can be shown that the `super-ageing' case 
$\mu>1$ is incompatible with basic requirements from probability theory \cite{Kurchan02}.
\bibitem{frust} It is conceivable that our finding of a distinct behaviour for 
$\eps=2$ might be a sort of borderline behaviour
when going from ferromagnets to frustrated magnet.
\end{thebibliography}
\end{document}